\documentclass[pra,twocolumn,superscriptaddress]{revtex4} 
\usepackage{epsfig}

\usepackage{graphicx}

\usepackage{graphicx}

\newcommand{\question}[1]{{}}

\newcommand{\tpaper}{this paper} 

\newcommand{\beq}{\begin{equation}}
\newcommand{\eeq}{\end{equation}}

\newcommand{\commentout}[1]{}

\begin{document}
\title{Supercurrents in an atom-molecule gas in an optical
ring lattice}
\author{T. Wang}
\affiliation{Department of Physics, University of Connecticut,
Storrs, CT 06269}
\author{J. Javanainen}
\affiliation{Department of Physics, University of Connecticut,
Storrs, CT 06269}

\author{S. F. Yelin}
\affiliation{Department of Physics, University of Connecticut,
Storrs, CT 06269} \affiliation{ITAMP, Harvard-Smithsonian Center
for Astrophysics, Cambridge, MA 02138}

\date{\today}
\begin{abstract}
Atom and molecule currents in a Fermi gas in the neighborhood of a
Feshbach resonance are studied in a one-dimensional optical ring
lattice by directly diagonalizing small models. A
rotational analogy of flux quantization is used to show that
fraction of the current is carried by particles with twice the mass
of an atom, which suggests pairing and superfluidity.

\end{abstract}

      \maketitle

Artificial condensed-matter materials constructed by loading
low-temperature atoms in an optical lattice hold promise as a
means to explore major outstanding questions such as the origin of
high-$T_c$ superconductivity, to realize what  heretofore have
been fundamental thought experiments in condensed matter physics,
and maybe to carry out quantum computations~\cite{LEW06,BUC05}. The
degree of control exerted by the experimenter in an optical lattice
stands in stark contrast to condensed-matter systems supplied by
Nature. For instance, the interactions between the atoms may be
controlled by varying a magnetic field around a Feshbach
resonance~\cite{STW76,TIE92,TIM98}.
Photoassociation~\cite{THO87,NAP94,WYN00} could
be similarly employed~\cite{DRU98,JAV99}.

In fact, though, Feshbach resonance and photoassociation entail
conversion of atom pairs into molecules, and vice
versa~\cite{KOH06}. In \tpaper\ we study flows in an
interconverting atom-molecule system in an optical lattice by
direct numerical solutions of small systems. In a close analogy
with past investigations of flux
quantization~\cite{SuperfluidityTJmodel}, we analyze a
one-dimensional optical lattice bent to a ring and rotated. We
find indications for one facet of superfluidity with fermions,
namely, that currents are carried by pairs of atoms reminiscent of
Cooper pairs. Complete pairing may be approached at the Feshbach
resonance when the atom-molecule coupling is strong, and when the
magnetic field is varied across the resonance the carriers of the
current  smoothly change between atoms and molecules.

What constitutes superfluidity is an interesting question in its
own right. In the experiments on superfluidity in optical
lattices~\cite{GRE02,SuperFermiOL} the criterion has been that,
once released from the lattice, the atoms from different sites
interfere. This, however, is an observation of off-diagonal
long-range order, not necessarily of superfluidity. A persistent
current is not a failsafe indicator of superfluidity either. In a
mesoscopic ring that holds electrons and is threaded by a
magnetic flux, the quantum mechanical ground state may carry a
finite current. From this viewpoint, experimental observations of
persistent currents in mesoscopic rings~\cite{MesoRing} are,
perhaps, not surprising. Instead, a frequently used criterion for
superfluidity in fermion systems is anomalous flux quantization.
Basically, in phenomena associated with the presence of the
magnetic field, the current
appears to
be carried
by particles
with twice
the charge
of an
individual
carrier~\cite{InsulatorMetalSuperCriteria,FluxQuantizationRing,
SuperfluidityTJmodel}.

Given that direct numerical solutions are feasible only in small
systems, studies of long-range off-diagonal order that would
model the experimental procedures~\cite{GRE02,SuperFermiOL} are
presently beyond our capabilities. We therefore look for anomalous
flux quantization. The remaining problem with neutral
atoms is that there is no direct coupling between magnetic field
and the center-of-mass motion. In principle one could use
laser-induced electromagnetic couplings to produce an effective
gauge field on the atoms~\cite{JUZ04,ZollerMagnetic,RealizeRing},
but here we envisage using rotation of the atomic sample as a
substitute for magnetic field.

Suppose that the prospective superfluid is tightly confined to a
torus with the circumference $L$, and that in the direction of the
torus, $x$, a rotated periodic potential $V(x-v t)$ is imposed,
where $v$ is the linear velocity along the torus.  When one
transforms the one-particle Schr\"odinger equations for the wave
function $\psi(x,t)$ to a co-moving frame, $\xi = x-vt$, $\tau=t$,
and defines $\psi(x,t) = e^{i m(v^2\tau/2+v\xi)/\hbar
}\Psi(\xi,\tau)$, in the transformed Schr\"odinger equation the
only remnant of the motion of the potential $V(x)$ is the twisting
boundary condition $\Psi(\xi+L,\tau)=e^{-i\Phi}\Psi(\xi,\tau)$,
where the rotation phase is $\Phi=mvL/\hbar$. The value of $\Phi$
only matters modulo $2\pi$.

Next assume that the optical-lattice potential is deep enough to
warrant a tight-binding approximation. The Wannier function for
each site $k=1,\ldots,N_s$, may be introduced as usual,
$u(x-x_k)$, but the annihilation operator of, say, a boson at site
$k$, $b_k$, is chosen in such a way that the corresponding wave
function is $e^{-ik\varphi}u(x-x_k)$, $\varphi = \Phi /N_s$ being
the phase change from one lattice site to the next. The purpose of
our convention is to revert the twisting boundary condition to the
usual periodic boundary condition. The cost is altered phases of
the hopping matrix elements, from a real $t$ to $t e^{\pm
i\varphi}$, in the ensuing Hubbard type model.

The same approach works even if multiple
interacting and interconverting species are present. We focus on a
model in which two types of fermionic atoms (annihilation
operators
$c_{k\uparrow}$ and $c_{k\downarrow}$) may combine on-site to a diatomic
bosonic molecule ($b_k$). The
Hamiltonian becomes
\begin{eqnarray}
{H\over\hbar} = & - &t_A\sum_{k,\sigma=\uparrow,\downarrow}
(e^{i\varphi_A}c^\dagger_{k+1,\sigma}c_{k,\sigma} + {\rm h.c.})\nonumber
\\
&-&t_M\sum_{k}
(e^{i\varphi_M}b^\dagger_{k+1}b_{k}+{\rm h.c.}).
\nonumber
\\    &+&\delta\sum_k b^\dagger_k b_k+g\sum_k
(b^\dagger_k\,c_{k \downarrow}c_{k\uparrow} + {\rm
h.c.})\,.\label{eq:Hamiltonian}
\end{eqnarray}
Here $g$ is the coupling strength for atom-molecule conversion and
$\delta$ is the detuning from the Feshbach resonance controlled by
the magnetic field~\cite{footnote1}. We have thereby formulated
what is known as a two-channel theory, with atoms and molecules as
separate albeit coupled degrees of freedom~\cite{KOH06}. We take
the masses of the $\uparrow$ and $\downarrow$ atoms to be equal,
whereupon the mass of the molecules is twice the mass of the atoms
and the two hopping phases satisfy $\varphi_M = 2\varphi_A$.
Different-mass atoms lead to incommensurate situations that do not
have magnetic-field counterparts, but we do not go there in this
paper. Note that Hamiltonian~(\ref{eq:Hamiltonian}) is the same as
that for a ring lattice threaded by a magnetic field
$B=2mvc/qL$~\cite{WuRotation}.

In the present model the conserved particle
number $\hat N = \sum_k \hat N_k$ is a sum over the lattice sites
of the local particle numbers $\hat N_k=c^\dagger_{k\uparrow}
c_{k\uparrow}+ c^\dagger_{k\downarrow}  c_{k\downarrow} + 2
b^\dagger_k b_k$.
$\hat N_k$ thus is a locally conserved quantity whose value only
changes as a result of transport. This gives the
identification of the operator for the current from the site $k$
to the site $k+1$ as
\begin{eqnarray}
\hat I_{k\rightarrow k+1} &=&
\sum_{\sigma=\uparrow,\downarrow}
it_A(e^{i\varphi_A}c^\dagger_{k+1,\sigma}c_{k,\sigma}-
e^{-i\varphi_A}c^\dagger_{k,\sigma}c_{k+1,\sigma})\nonumber\\
&+&2 it_M(e^{i\varphi_M}b^\dagger_{k+1}b_k-
e^{-i\varphi_M}b^\dagger_kb_{k+1})\,.
\label{TOTCUR}
\end{eqnarray}
If the system is in an energy eigenstate in which
there is a steady flow of atoms and/or molecules along the
lattice, the value of the current $I\equiv \langle I_{k\rightarrow
k+1}\rangle$  and the energy of the state
$E(\Phi_A,\Phi_M)$ regarded as a function of the phase parameters
satisfies
\begin{equation}
\label{eq:CurrentEnergy} I =
-\frac{1}{\hbar}\left(\frac{\partial
E}{\partial\Phi_A}+2\frac{\partial E}{\partial\Phi_M}
\right).
\label{CUREQ}
\end{equation}
Since $E$ is a period function of $\Phi$ with period $2\pi$, so
does $I$ as shown by Eq.~(\ref{CUREQ})~\cite{BYE61}. Here we may
also compute the current numerically from Eq.~(\ref{TOTCUR}) and
easily separate it to atomic and molecular components, but
Eq.~(\ref{CUREQ}) is convenient for qualitative discussions.

A way to compare the results for different system sizes is needed
in order to ascertain that the outcome is not determined by
limitations of the numerics. To this end we first note that for a
nonrotating lattice, $v=0$ and $\varphi_{A,M}=0$, the ground state
is obviously translationally invariant and possesses a left-right
symmetry: $\langle
c^\dagger_{k+1,\uparrow}c_{k\uparrow}\rangle=\langle
c^\dagger_{k'+1,\uparrow}c_{k'\uparrow}\rangle$ and $\langle
c^\dagger_{k+1,\uparrow}c_{k\uparrow}\rangle=\langle
c^\dagger_{k-1,\uparrow}c_{k\uparrow}\rangle$ for all $k$ and
$k'$. The expectation values $\langle
c^\dagger_{k+1,\uparrow}c_{k\uparrow}\rangle$ are therefore real
at $v=0$. The motion of the lattice enters only through the phase
parameters $\varphi_{A,M}$. Assuming that perturbation theory in
$\varphi_{A,M}$ is valid, Eq.~(\ref{TOTCUR}) shows that in the
limit $v\simeq0$ or $\varphi_{A,M}\simeq0$ the current is a linear
function of $\varphi_{A,M}=\Phi_{A,M}/N_s$. On the other hand, a
sensible thermodynamic limit should be taken so that the number of
lattice sites $N_s$ and the invariant particle number $N$ both
tend to infinity in such a way that the average occupation number
per site $N/N_s$ remains a constant. Ground-state expectation
values such as $\langle
c^\dagger_{k+1,\uparrow}c_{k\uparrow}\rangle$  may be viewed as
some functions $h_{\uparrow\uparrow}(N_s,N/N_s)$, and in the limit
$N_s\rightarrow\infty$ with $N/N_s$ = constant they evidently tend
to constants that depend only on the fixed ratio $N/N_s$. Putting
these considerations together, we see that, for given total phases
$\Phi_A$ and $\Phi_M$, the only possible $v$-independent scaling
of the current that leads to a nonzero and finite current in the
thermodynamic limit is $I_T = N_s I$.

In this paper direct diagonalization based on the Lanczos
algorithm~\cite{ARPACK} is employed to find the eigenvalue
(energy) and eigenvector of the ground state of the
Hamiltonian~(\ref{eq:Hamiltonian}), and the results are then used
to calculate various observables at zero temperature. In this
approach arbitrary parameter values may be accommodated without
approximations, although only for small numbers of atoms and
lattice sites. From now on we  use the value of the atomic
tunneling matrix element $t_A$ as the frequency scale. The number
of $\uparrow$ and $\downarrow$ fermions is always the same.  Until
further notice we also assume that the molecules do not tunnel,
whereupon $t_M=0$ and the parameters $\varphi_M$ and $\Phi_M$ do
not enter at all.

\begin{figure}[t]
       \centerline{\includegraphics[clip,width=.9\linewidth]{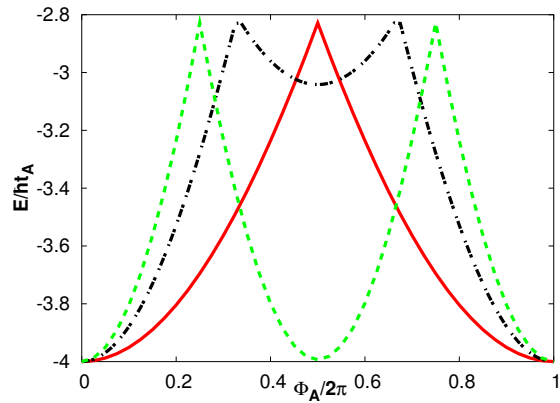}}
        \caption{\protect\label{DOUBLING} (Color online) Ground state
energy $E$ as
a function of the rotation phase $\Phi_A$ for $g=0$ (solid line),
$g/t_A=5$ (dash-dotted line), and $g/t_A=100$ (dashed line). The
energies corresponding to $g/t_A=5$ and $g/t_A=100$ have been scaled
linearly so that the ranges of $E$ coincide for all three $g$. The
fixed parameters are $\delta=0$, $N_s=4$, $N=2$.}
\end{figure}

We first study the resonance, $\delta=0$, varying the coupling
strength $g$. The dependence of $E(\Phi_A)$ on the phase $\Phi_A$
is the weaker the larger is the coupling $g$. The current $I$
therefore decreases with $g$. With this in mind, we plot in
Fig.~\ref{DOUBLING} the energies $E$ as a function of $\Phi_A$ for
the coupling strengths $g/t_A=0,5$, and 100, scaling the energy
linearly in such a way that the minima and the maxima for each $g$
coincide. The invariant particle number is $N=2$, and the number
of lattice sites is $N_s=4$.

The main observation is that with increasing $g$ the
energy changes from being a $2\pi$ periodic function of $\Phi_A$ to
having the period
$\pi$. The current, basically the derivative $dE/d\Phi_A$, behaves as
if it were a sum of two components corresponding to the carrier
masses $m$ and $2m$, with the mass $2m$ taking over as the coupling
strength is increased. In analogy with  the earlier studies
of the influence of the magnetic
flux~\cite{InsulatorMetalSuperCriteria,FluxQuantizationRing,
SuperfluidityTJmodel}, we interpret this to mean that some of the
fermions are paired, suggesting superfluidity. Moreover, the fraction
of fermions appearing as pairs increases with increasing coupling
strength.

\begin{figure}[t]
\centerline{\includegraphics[clip,width=.9\linewidth]{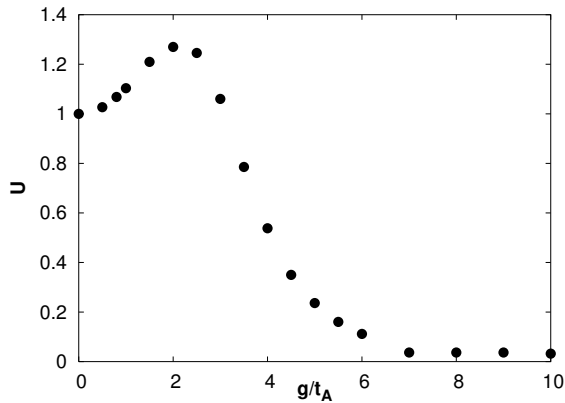}}
        \caption{\protect\label{ATMOLSWITCH}  (Color online) Unpaired
fraction $U(g)$,
a measure of the fraction of the current that is due to non-paired
fermions, as a function of the coupling strength $g$.  The fixed
parameters are $\delta=0$, $N_s=8$, $N=2$.  }
\end{figure}

We next quantify the fraction of the current that is due to unpaired
fermions. One might think that Fourier analysis of
$E(\Phi_A)$ gives straightforward answers, but there are
problems. For instance, even with $g=0$ and no pairing at all,
there is still a Fourier component in $E(\Phi_A)$ with the period
$\pi$. We therefore use the ratio of the Fourier component of the
current with the period
$2\pi$ to an estimate of the current,
normalized so that the value at $g=0$ equals unity, as the measure of
the unpaired fraction. Specifically, we first find
\begin{equation}
f(g) = \frac{|\int_{0}^{2\pi} d\Phi\, \sin\Phi\,
I(\Phi;g)|}{\max_{\Phi}[I(\Phi;g)]-\min_\Phi [I(\Phi;g)]},
\end{equation}
and define $U(g)=f(g)/f(0)$. A representative unpaired fraction $U$ is plotted
in Fig.~\ref{ATMOLSWITCH} as a function of the coupling strength
$g$, for $\delta=0$, $N=2, N_s=8$. Unpaired fractions larger than
unity are an unfortunate artifact of our definition, but the
qualitative result is clear: With an increasing coupling strength
the fraction of unpaired current carriers tends to zero.

\begin{figure}[t]
\centerline{\includegraphics[clip,width=.9\linewidth]{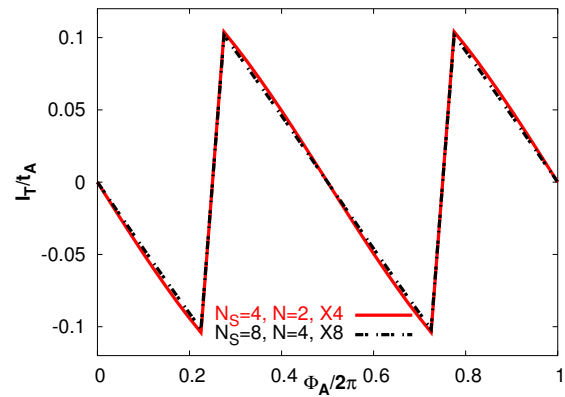}}
        \caption{\protect\label{SCALING} (Color online) Scaled current as
a function
        of the rotation phase for lattices with $N_s=4$, $N=2$ (solid
line) and  $N_s=8$, $N=4$ (dash-dotted line). The actual currents
have been multiplied by the factors indicated in the legend. This
figure is for Feshbach resonance, $\delta=0$, and the
atom-molecule coupling  $g/t_A=100$ ensures that the fermions are
almost completely paired.}
\end{figure}

To investigate the effects of the necessarily small size of the
system on the results we draw in Fig.~\ref{SCALING} the current
scaled by the number of lattice sites, $I_T(\Phi_A)=N_s
I(\Phi_A)$, as a function of the rotation phase $ \Phi_A$ for the
system sizes $N_s=4$ and $N_s=8$, both with $N/N_s=1/2$. We set
$\delta=0$, and choose $g/t_A=100$. If $I_T$ were an invariant in
the thermodynamic limit, in the limit of large $N_s$ such curves
should overlap. They do so quite well already in these small
systems. Our interpretation is that, suitably scaled, the results
for $N_s=8$ and $N=4$ already represent well the physics of the
limit of a large number of lattice sites.

We have also looked into the effect of varying the atom number $N$
for a fixed number of sites $N_s$ on the fermion pairing behavior.
In our numerics there is not much room for variation in $N$, and in a
small system even a single unpaired fermion may make a major difference.
With these caveats, in the strong-coupling case $g\gg t_A$ we have not
detected a qualitatively significant effect of atom number on the
fraction of paired fermions as long as the number of fermions is even.

In our final example we turn to the variation of the current with
the detuning $\delta$, the situation of BEC-BCS crossover where
the system at near-zero temperature equilibrium is expected to
change from an ideal Fermi gas to a BCS superfluid and then to a
BEC of the diatomic molecules when the detuning is scanned across
the resonance starting from a large positive value. In such a
process the increasing fraction of molecules, the ``bare''
molecules denoted by $b_k$ in our theory, has been observed
experimentally~\cite{PAR05} in a nonlattice gas, and successful
comparisons of two-channel mean-field theories with the
experiments have been reported~\cite{CHE05,ROM05,JAV05}. Our model
automatically leads to the period $\pi$ for the molecular currents
and as such says nothing about the superfluidity of the molecules,
so that molecular currents are somewhat off the theme of the
paper. Nonetheless, we shall demonstrate the change in the nature
of the current carriers from atoms to molecules in a lattice upon
the crossover.

\begin{figure}[ht]
\centerline{\includegraphics[clip,width=.9\linewidth]{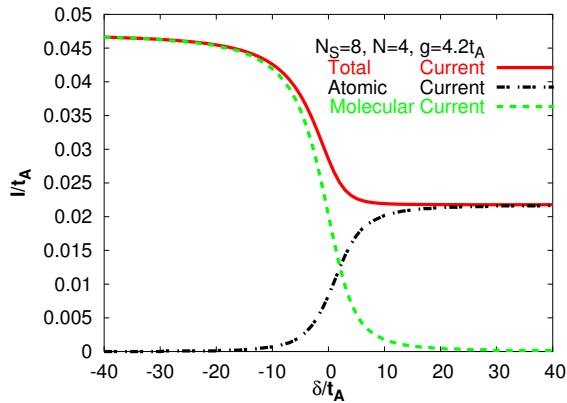}}
        \caption{\protect\label{XO} (Color online) Total current (solid
line), atomic current (dash-dotted line) and molecular current (dashed
line) as a function of detuning in a lattice with $N_s=8$, $N=4$,
$t_M/t_A=1$, and $g/t_A=4.2$.  }
\end{figure}

In order to allow for molecular currents in the first place we
need to choose a nonzero hopping matrix element $t_M$, thus we set
$t_M=t_A$, and pick $g/t_A=4.2$ so that about half of the
current is expected to be paired according to
Fig.~\ref{ATMOLSWITCH}. We plot in Fig.~\ref{XO} both the total
current  and the atomic and molecular currents separately as a
function of the varying detuning $\delta$. The system size is
$N_s=8$, $N=4$. The expected smooth transition between atomic and
molecular currents is observed.

Although our study is meant to be an in-principle theoretical
discussion, a few comments about experiments are in order.
Toroidal traps have been realized already~\cite{GUP05}, and there
is a proposal to make an optical ring lattice~\cite{RealizeRing}.
As usual, the lattice potential results from interference between
two light beams. Also as usual, the lattice potential may be made
to move simply by taking the two beams to have different
frequencies, here possibly by means of an optical delay line
incorporating a (slowly) moving mirror. Analysis of the velocity
distribution of the atoms has been discussed in
Ref.~\cite{RealizeRing}. Next, the coupling strength for the usual
834$\,$G Feshbach resonance in ${}^6$Li is large compared to the
widths of the energy bands in any conceivable optical lattice.
This implies that a single-band model such
as~(\ref{eq:Hamiltonian}) is inadequate~\cite{DIE06}. A faithful
realization of the Hamiltonian~(\ref{eq:Hamiltonian}) may have to
await further developments in photoassociation~\cite{WIN05}, which
in principle permits a full control of the coupling strength. As
the tunneling matrix elements are exponentially small in the mass
of the particle, the case with $t_A\gg t_M$ is probably the
experimental default. However, tunneling matrix elements can in
principle  be controlled with light by using optical
transitions~\cite{RUO02}. Overall, numerous and severe challenges
must be overcome before our schemes are amenable to experiments,
but the obstacles seem to be a matter of technology.

In sum, we have studied currents in an interconverting atom-molecule
gas in an optical ring lattice by direct numerical solutions of
small two-channel models. The center-of-mass motion of a neutral
atom or molecule does not directly couple to the magnetic field, so we
envisage rotation of the ring as a substitute to magnetic
field for realizing an analog of flux quantization
in a superconductor. We find flux quantization as if part of the
current were carried by atom pairs, which constitutes indirect
evidence for superfluidity. With an
increasing atom-molecule coupling, an increasing fraction of the
current is carried by atoms pairs. In analogy with the BEC-BCS
crossover, variation of the magnetic field in the neighborhood of the
Feshbach resonance leads to a smooth switching of the current
carriers from atoms to molecules. The experimental challenges with
our schemes are severe, but appear to be purely technical.

This work is supported by Innovation
Grant from the Research Corporation, and by NSF (PHY-0354599) and
NASA (NAG3-2880).


\begin{thebibliography}{16}
\bibitem{footnote1} Confining the  atoms to a
lattice shifts the Feshbach resonance, so that the detuning
$\delta$ has no simple relation to the detuning that applies in a
(dilute) free gas for a given magnetic field~\protect\cite{DIE06}.
Besides, the nature of the Feshbach resonance is materially
different in one and three spatial dimensions~\protect\cite{MAC07}.
For the present purposes we use
$\delta=0$ as a qualitative indicator of the
Feshbach resonance.
\end{thebibliography}
\end{document}